# Comment on "Physical Origin and Generic Control of Magnonic Band Gaps of Dipole-Exchange Spin Waves in Width-Modulated Nanostrip Waveguides"

K. Di,[1] H. S. Lim,[1,*] V. L. Zhang,[1] M. H. Kuok,[1] S. C. Ng,[1] M. G. Cottam,[2] H. T. Nguyen[2]

[1]*Department of Physics, National University of Singapore, Singapore 117542*
[2]*University of Western Ontario, Department of Physics and Astronomy, London, Ontario N6A 3K7, Canada*

In Ref. [1] Lee *et al*. reported the existence of large magnonic bandgaps in one-dimensional width-modulated Permalloy nanostripe waveguides based on OOMMF simulations [2]. However, as the symmetry of the magnetic field pulse they applied to excite the spin waves (SWs) was not general, the entire set of SW branches with A symmetry was omitted from the magnonic band structures (see below). This omission has unfortunately led to misleading conclusions of, for instance, the number, width and position of the bandgaps. We present here the full band structure based on three different theoretical approaches that gave consistent predictions, thus corroborating the methods employed, namely, a microscopic approach, OOMMF simulations, and a method based on the linearized Landau-Lifshitz equation. Further, we provide a physical interpretation using group theory.

The dispersion relation of the [$P_1$, $P_2$] = [9nm, 9nm] waveguide was obtained by numerically solving the linearized Landau-Lifshitz equation for a unit cell with the Bloch–Floquet boundary conditions applied. In the microscopic calculation, based on [3], the waveguide structure was represented in terms of a large number of cubic cells, each with an effective spin, and a Hamiltonian approach was adopted for the interacting spin system including the exchange and dipole-dipole terms. Discrete-lattice dipole sums were performed, instead of utilizing Maxwell's equations, with the cell size being less than the exchange length. Both these methods do not depend on using an excitation field.

Interestingly, the dispersion curves (Fig. 1a) contain branches, labeled 2 and 4, not reported in [1]. The presence of these additional modes, which have A symmetry magnetization profiles (Fig. 1b), drastically reduces the first bandgap width from the reported 11 to 3 GHz and the second bandgap from 16 to 2.2 GHz. As the magnetization is an axial vector, the symmetry of the magnetic ground state is the $C_{2h}$ group [4]. The groups of the wavevectors at the Γ, Δ (a general point in the first Brillouin zone) and X points are $C_{2h}$, $C_2$, and $C_{2h}$, respectively. The symmetry assignment of the modes, based on their profiles (Fig. 1b) of the out-of-plane component $m_z$ of the dynamic magnetization, is presented in Fig. 1a.

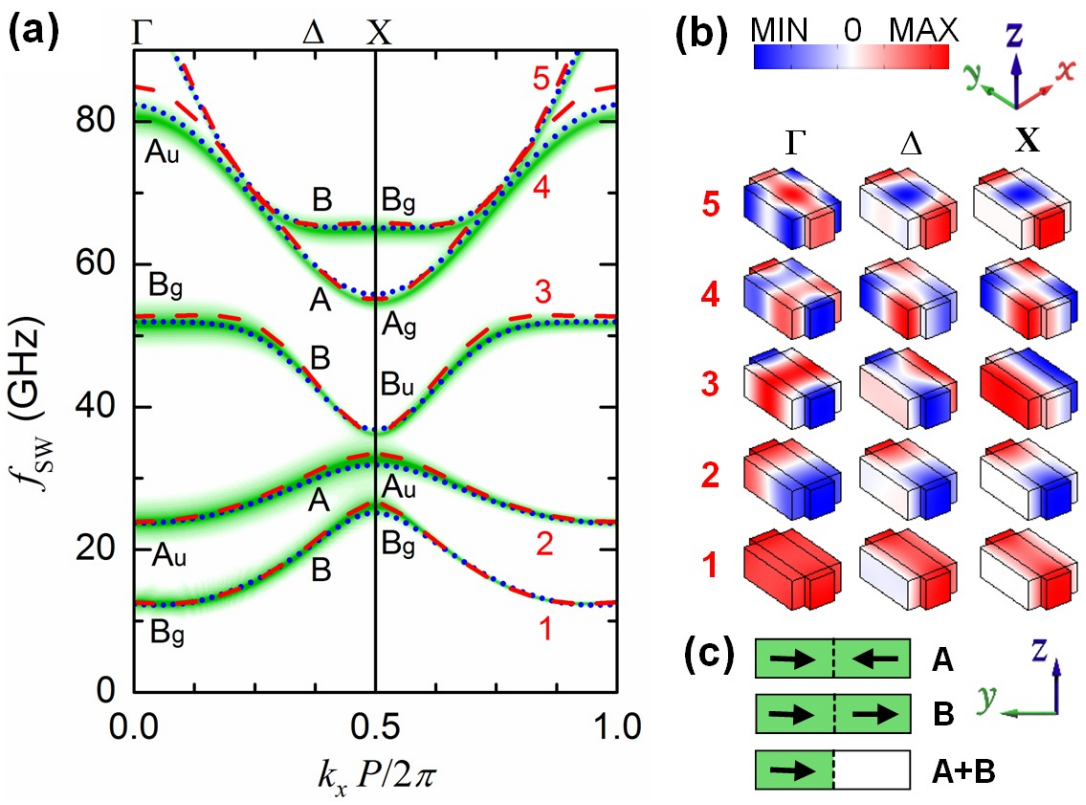

FIG. 1 (color online). (a) Dispersion relations for the waveguide obtained from linearized Landau-Lifshitz equation method (red dashed lines), microscopic calculations (blue dotted lines) and OOMMF simulations (green lines). The symmetry assignment of the modes is indicated. (b) Mode profiles of $m_z$ for SW branches (labeled with 1-5) at Γ, Δ and X points. (c) Instantaneous spatial profiles ($y$-$z$ cross-section) of the excitation field, with different symmetries (A, B, and both). The shaded area denotes the section of the waveguide where the field is applied.

In our OOMMF simulations, based on a cell size of 1.5×1.5×1.5 $nm^3$, a sinc magnetic field pulse was applied in the $y$-direction to a 1.5×15×10 $nm^3$ central section of a 4μm-long waveguide (shaded area of bottom panel of Fig. 1c). As the field can be expressed as a linear combination of the basis functions that transform as the A and B irreducible representations of the $C_2$ group, it can therefore simultaneously excite modes of both A and B symmetries. The dispersion curves, obtained by performing the Fourier transform of $m_z$ in space and time (summing contributions from all cells), are included in Fig. 1a. Further calculations confirm that only A (B) symmetry modes can be excited by an excitation field of A (B) symmetry (Fig. 1c). Hence, the use of an excitation field of B symmetry by Lee *et al.* precluded their observation of A symmetry modes [1,5] and led to misleading conclusions for the bandgaps. Consequently we find that the reported numbers of bandgaps of most of the waveguides studied are wrong. For instance, the [$P_1$, $P_2$] = [15nm, 15nm] structure has only three rather than the stated five bandgaps for the same frequency range, as well as significantly narrower bandgap widths than claimed.

[*] Electronic mail: phylimhs@nus.edu.sg.

This work was funded by the Ministry of Education, Singapore under grant R144-000-282-112.

REFERENCES

[1] K.-S. Lee, D.-S. Han, and S.-K. Kim, Phys. Rev. Lett. **102**, 127202 (2009).
[2] see http://math.nist.gov/oommf. OOMMF User's Guide Version 1.2a4.
[3] H. T. Nguyen and M. G. Cottam, J. Phys. D: Appl. Phys. **44**, 315001 (2011); see also H. T. Nguyen and M. G. Cottam, J. Appl. Phys. **111**, 07D122 (2012).
[4] M. S. Dresselhaus, G. Dresselhaus, A. Jorio, *Group Theory: Application to the Physics of Condensed Matter* (Springer, Berlin, 2008).
[5] S.-K. Kim, J. Phys. D: Appl. Phys. **43**, 264004 (2010).